# Title

The First French COVID19 Lockdown Twitter Dataset

# Authors

Sophie Balech[1]
Christophe Benavent[2]
Mihai Calciu[3]

# Affiliations

1. IAE Amiens, University of Picardie, France
2. Department of Management Science, Paris-Nanterre University, France
3. IAE Lille, Lille University, France

# Abstract

In this paper, we present a mainly French coronavirus Twitter dataset that we have been continuously collecting since lockdown restrictions have been enacted in France (in March 17, 2020). We offer our datasets and sentiment analysis annotations to the research community at https://github.com/calciu/COVID19-LockdownFr. They have been obtained using high performance computing (HPC) capabilities of our university's datacenter. We think that our contribution can facilitate analysis of online conversation dynamics reflecting people sentiments when facing severe home confinement restrictions determined by the outbreak of this world wide epidemic.
We hope that our contribution will help decode shared experience and mood but also test the sensitivity of sentiment measurement instruments and incite the development of new instruments, methods and approaches.

# Introduction

This paper helps explore the value of social data for the study of human phenomena among which the covid19 epidemic is obviously a major one. Far beyond health aspects its consequences are social, political and economic. The way people experience it, at least in France, comes from an emergency regime, widely adopted across countries, which aims to keep the majority at home.
The feeling experienced is that of confinement (lockdown). A recommendation that has become an obligation supported by a regulatory system that controls, nudges and sanctions, a hopefully temporary deprivation of liberty. A social life limited to the home and some electronic means of communication.
A major network where states of mind can express themselves freely is Twitter. Very quickly there emerged a daily confinement hashtag (in french #ConfinementJour). It helps share experience and mood. We use this channel in order to test the sensitivity of sentiments and emotions measurement instruments, and, hopefully, better understand human reactions to a brutal anthropological shock (the clash of an advanced and mobile society with a natural contingency - a virus which circulates from mouth to mouth and travels in business class). We tried to record how people experience this shock day after day, by capturing social chatter and measuring feelings, emotions and concerns.

# Data collection

This dataset includes data from the publicly available Twitter Rest API. The daily gathering process continued since March 17 until at least May 11, when lockdown restrictions in France shall be loosened. Keywords used were #ConfinementJourXx hashtags. The Xx indicator means the number of days from the beginning of the lockdown policy, and is incremented each day. Note that the collection had to observe the free Rest API limitations regarding tweets extraction.

Due to the particularity of the chosen hashtag, most of the tweets are expressed in french. While incrementing each day the final dataset, we filter the new data to avoid duplicated tweets.

Table 1 - Number of #ConfinementJourXx tweets in this dataset

| ISO | Language | No. tweets | % total tweets |
|---|---|---|---|
| fr | French | 2474086 | 95.22 |
| en | English | 31614 | 1.22 |
| es | Spanish | 2997 | 0.12 |
| ca | Catalan | 2412 | 0.09 |
| pt | Portuguese | 1817 | 0.07 |
| it | Italian | 1153 | 0.04 |
| ht | Haitian | 953 | 0.04 |
| ja | Japanese | 731 | 0.03 |
| tr | Turkish | 672 | 0.03 |
| in | Indonesian | 588 | 0.02 |
| de | German | 505 | 0.02 |
| no | Norvegian | 345 | 0.01 |
| eu | Tagalog (Philipnes) | 280 | 0.01 |
| tl | Arabic | 264 | 0.01 |
| ar | Basque | 263 | 0.01 |
| ro | Romanian | 254 | 0.01 |
| et | Dutch | 249 | 0.01 |
| nl | Estonian | 245 | 0.01 |
| da | Danish | 192 | 0.01 |
| sv | Swedish | 185 | 0.01 |

We make a distinction between full and clean versions of the dataset. The full dataset contains all tweets while the clean version eliminates retweets. We choose to create a clean dataset with no retweets for later NLP analysis.

Table 2 - Number of #ConfinementJourXx tweets per week

| Week | Tweets |
|---|---|
| Week 1 (17.03-22.03) | 516216 |
| Week 2 (23.03-29.03) | 510611 |
| Week 3 (30.03-05.04) | 373317 |
| Week 4 (06.04-12.04) | 297451 |
| Week 5 (13.04-19.04) | 211096 |
| Week 6 (20.04-26.04) | 262207 |
| Week 7 (27.04-03.05) | 226200 |
| Week 8 (04.05-10.05) | 199942 |
| Total | 2598249 |

The datasets were used to create an annotated corpus using a dictionary approach with different emotion and sentiment annotators. Figure 1 outlines the steps taken to build our datasets.

Figure 1 - Dataset gathering and construction steps

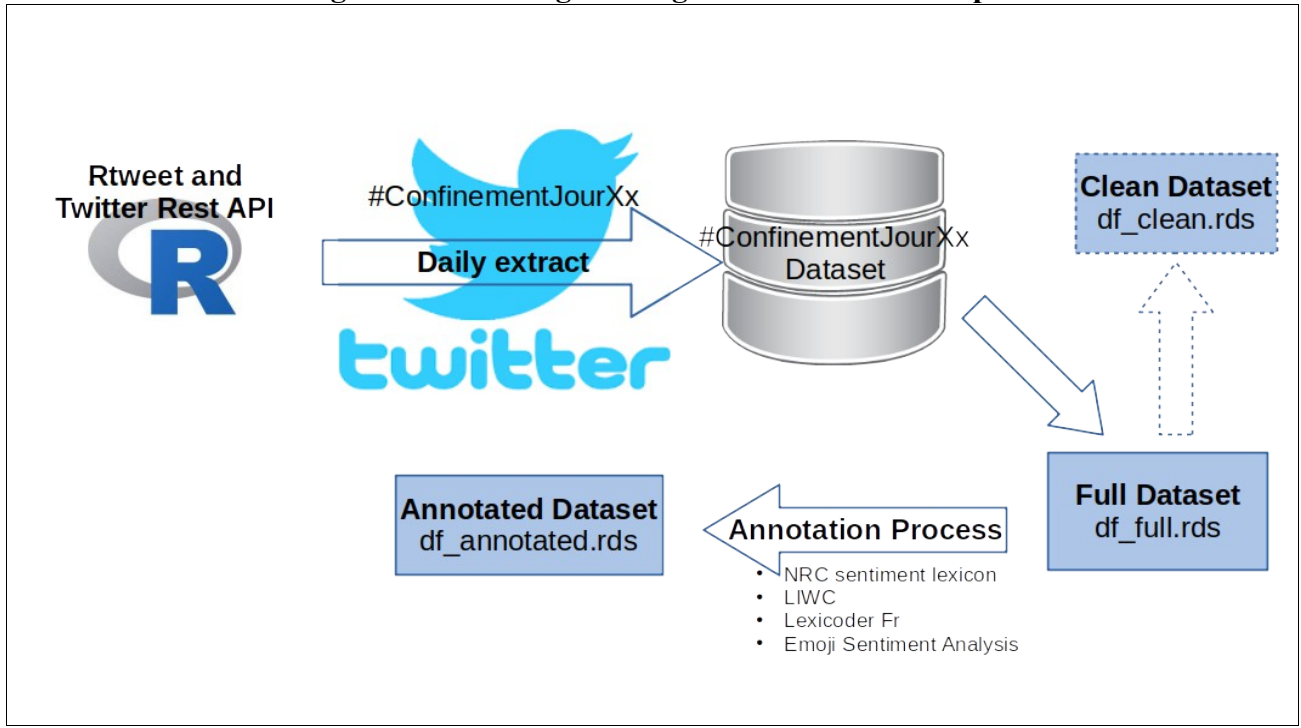

As shown in figure 1, our work environment is essentially based on the R language and environment for statistical computing. We use the rtweet package (Kearney,2019) combined with Twitter Rest API to extract on a daily basis all tweets mentioning #ConfinementJourXx, with Xx for the current day since the start of lockdown policy in France. We aggregate the daily records into a full dataset, *df_full.rds*. We filter this dataset to keep only original tweets and save it as the clean dataset, *df_clean.rds*.

## Data Annotations

The annotation process consisted in extracting from each tweet text, words expressing sentiments and emotions and counting their frequency. For this purpose we used several dictionaries and lexicons:

- The NRC (National Research Council Canada) emotion lexicon (Mohammad and Turney, 2013): based on the Putchnik (1980) scale of sentiments and emotions and integrated to R's suyzhet package (Jockers, 2015) served to score each tweet on sentiments (positive or negative) and emotions (anger, fear, anticipation, trust, surprise, sadness, joy, and disgust) based on word occurrence.

- The Linguistic Inquiry and Word Count (LIWC) system (Pennebaker & al., 2015) that counts the percentage of words that reflect different emotions, thinking styles, social concerns, and even parts of speech, has been applied using the quatenda.dictionaries[1] R package (Benoit, 2018).

---

1  https://github.com/kbenoit/quanteda.dictionaries

- The Lexicoder Sentiment Dictionary (Young & Soroka, 2012) with its LSD french version (Duval & Pétry, 2016) has been also used with quanteda in order to count terms expressing positive and negative sentiments.

- The Emoji Sentiment Ranking (Kralj Novak, 2015) has been applied to associate sentiment scores to all emojis present in a tweet and retain a minimum, average and maximum score for those emojis.

This annotation process became soon the computational bottleneck of our approach. After collecting, more than two million tweets, the annotation lasted more than ten hours. We had to find a method to accelerate computations by splitting the tweets into chunks and applying these calculations in parallel on each chunk. We called this method "GRAPPA" (GeneRal Approach for Parallel Processing Annotations) it uses R's "Map" and "Reduce" functions. First we adopted implicit or multicore parallelism on one workstation with an 8 core CPU (Intel I9 microprocessor) and 16G memory (RAM) using R's "parallel" package (R Core Team, 2020) with its "mcapply" function that allowed for a two to four fold time reduction. As this was not enough we switched to cluster parallelism using the authors' universities resources and R's rslurm package (Marchand & Carroll, 2019) that submits R calculations to the cluster's job manager SLURM (Simple Linux Utility for Resource Management). The result was stunning: computation time fell to less than ten minutes, when using 16 nodes with 2 cores each. This huge difference in computing time between multicore and cluster parallelism is due to the fact that in multicore parallelism the available cores share the same memory while in cluster parallelism each node has its own memory..

The annotations resulted in several distinct datasets (*df_nrcliwc.rds, df_lsd.rds and df_emojis.rds*) and are freely available for research.

In compliance with Twitter's Terms & Conditions, collected tweet texts cannot be released. We release therefore only tweet IDs that researchers can use to retrieve the full tweet objects using so called "hydrating" techniques. In addition to the collected datasets containing only tweet identifiers we offer three annotations datasets containing emotions and sentiments frequencies and/or scores. We continue to update our datasets on a weekly basis, until the end of the lockdown or the tweets stream.

# Data Records

The datasets are made available at https://github.com/calciu/COVID19-LockdownFr.
Sentiment and Emotion Annotations datasets (in .csv format) are available in the "LockdownAnnot" folder (see table 3). Their names have the following pattern dff_method_numberOfDays_chunk.csv (dff meaning data frame free and the method is nrc, lsd or emos)

**Table 3 - Tweet Sentiment and Emotion Annotations datasets**

| Annotations File Names | Description | Example |
|---|---|---|
| dff_nrc_55_1.csv<br>dff_nrc_55_2.csv | NRC files contain ten columns the first eight represent emotions frequencies and the last two sentiments frequencies per tweet | anger,anticipation,disgust,fear,joy,sadness,surprise,trust,negative,positive<br>1,1,0,2,1,1,1,4,2,2 |
| dff_lsd_55_1.csv | LSD files contain four columns | Neg_lsdfr,Pos_lsdfr,nb,id |

| dff_lsd_55_2.csv | negative and positive sentiment frequency, number of words and identification number of the tweets | 0,0,1,1<br>0,0,20,2<br>0,1,38,3<br>1,1,26,4 |
|---|---|---|
| dff_emos_55_1.csv<br>dff_emos_55_2.csv<br>dff_emos_55_3.csv | Emoji files contain five columns. The first is a list of emojis separated by ";", the second indicates the number of emojis per tweet. The other three indicate the minimum, average and maximum sentiment score (on a scale from -1 to 1) | emoji,nemoji,min_sentsc,mean_sentsc,max_sentsc<br>;,0,NA,NA,NA<br>Ч ;Ч ;Ч ;,3,0.221,0.221,0.221<br>ȳ ;,1,0.43,0.43,0.43<br>Ӎ ;Ч ;ḍ ;,3,0.178,0.2935,0.409 |

The Tweet-IDs that help recover (hydrate) all collected datasets are organized as follows:

- Tweet-ID files are stored in two folders LockdownDays and LockdownOther (see table 4). The first has tweets collected with the containment day hashtag #ConfinementJourX, the second one contains tweets collected using a larger set of hashtags indicating the containment period.
- The file names have the following pattern: a prefix "df_ids" followed by either "jX" indicating the day (jour) for the first folder and by "confX" meaning confinement (lockdown) for the second.

**Table 4 -Tweet IDs datasets**

| TweetIDs File Names | Description | Example |
|---|---|---|
| df_idsj1-8.txt<br>df_idsj9.txt<br>…<br>df_idsj55.txt | Text files representing lockdown days with one tweet ID per row | 1242820357830123520<br>1242813166012293120<br>1242810902178664452<br>1242798340095475712 |
| df_idsconf1_1.txt<br>df_idsconf1_2.txt<br>df_idsconf2.txt | Text files representing a larger set of hashtags concerning the lockdown period with one tweet ID per row | 1245748350567297024<br>1245748349371920389<br>1245748343785115648 |
| df_twdata.rds | Date and time meta-data elements | 2020-03-25 14:27:18 |

The tweet ids can be hydrated using several easy to use tools that have been developed for such purposes like Social Media Mining Toolkit (Tekumalla & Banda,2020), Hydrator[2] or the python language module called twarc[3]. The deliverables include besides tweet ids, code to hydrate the tweets. We provided the date and time meta-data elements on our dataset for groups not interested in the tweet's text but only in the available annotations and for those wanting to target their research questions to certain days to avoid having to hydrate the whole resource at once.

We also welcome any additional data that provide new tweets to our resource or additional methods and code to further investigate such data.

---

2  https://github.com/DocNow/Hydrator
3  https://github.com/DocNow/twarc

# Code Availability

All the code used to extract, preprocess, annotate and analyze tweets is available on a public github repository (https://github.com/BenaventC/BarometreConfinement). The code is freely available and the dataset is licensed under Creative Commons Non Commercial Attribution-ShareAlike 4.0 International (CC BY-NC-SA 4.0) License, and should only be used for research purposes. In order to use the rtweet library and the Twitter Rest API, a Twitter developer account is required.

# Acknowledgements

This work is partially supported by the COMUE Paris-Lumière.

# Competing Interest

The authors of this paper have no competing interest.